# On the problems of creating a nuclear-optical frequency standard based on $^{229}$Th

F. F. Karpeshin, L. F. Vitushkin

D.I.Mendeleyev Institute for Metrology
VNIIM, Saint-Petersburg, Russia

## Abstract

The most probable candidate for the role of a nuclear optical standard is the 8.2-eV isomer of the 229mTh isotope of the thorium nucleus. Ways of using the resonant properties of the electron shell as an optical resonator to create laser-nuclear technologies necessary for the optical pumping of nuclear isomers and other manipulations of atomic nuclei leading to the creation of a next-generation frequency standard and nuclear-optical clocks based on them are discussed. Deep relations between the physics of resonance electron-nuclear interactions and the true solution of the thorium puzzle are shown.

The implementation of the project requires the refinement of the energy of the isomer up to the width of the nuclear line. This can be done by resonant optical pumping. Its practical implementation is impossible without using the resonant properties of the electron shell. In neutral atoms, they are reduced to internal conversion; in ionized atoms, resonant conversion takes its place. The first path is supposed to be implemented on the basis of the JILA university. Internal conversion leads to a broadening of the nuclear line by nine orders of magnitude, which facilitates the search for resonance to a practically feasible level.

The article discusses the second way in detail. It is shown that it will increase the efficiency of the experiment by another 3–5 orders of magnitude. The article discusses important principles of resonant optical pumping, such as the presence of a finite width in the intermediate electronic state, and others that are usually overlooked with a fatal result for the experiment. The wide application of the various physics of these processes will predetermine a revolutionary leap in the development of new laser-nuclear technologies.

*Contacts: Karpeshin Feodor ([fkarpeshin@gmail.com](mailto:fkarpeshin@gmail.com)).*

**Introduction**

Good prospects for creating a next-generation frequency standard and clock are promised by the use of spectral lines corresponding to nuclear isomer transitions. Nuclei, being located in the center of the electron shell, are weaker compared to atomic or molecular systems, subject to the effects of external and intracrystalline fields. These lines are narrow and stable. The problem is that transitions in most nuclei have energies of tens of keV. It is problematic to manipulate such transitions with the help of lasers. There is a unique nuclide $^{229}$Th whose excited state is located at a height of $\omega_n$ = 8.338(24) eV (Kraemer, 2022). The observed lifetime in neutral atoms turned out to be 10 µs. However, the intrinsic lifetime of the nucleus is much shorter. The point is that in neutral atoms the decay of the isomer is greatly enhanced due to the internal conversion (IC) channel. IC becomes the dominant decay channel with ICC (internal conversion coefficient) $\alpha(M1) = 10^9$. Accounting for the IC leads to an increase in the natural width of the isomeric line from $\Gamma_n$ = 0.844×10$^{-19}$ eV (10$^{-5}$ Hz) to $\Gamma_a$ = 0.8×10$^{-10}$ eV (10 kHz).

So far, there is no necessary condition for designing a frequency standard: the transition energy is unknown with the required accuracy. To determine it, it is planned to use the IC channel of isomer decay in neutral thorium atoms (Von der Wense, 2016). At one time, it was the IC that made it possible to record the first experimental observation of isomer decay (Seiferle, 2021). To excite the isomer, it would be ideal to use a VUV tunable laser near the resonant frequency. By tuning the frequency, the energy of laser photons is scanned. When resonance is reached during scanning, the isomer is excited by absorption of a laser photon. Then the isomeric state of the nucleus decays into the ground state by way of IC. In turn, conversion electrons are registered by a special multichannel detector. Thus, the detection of a signal from a multichannel detector should indicate that the isomeric transition frequency is equal to the laser frequency. However, there are practically no such lasers with a wavelength in the range of about 150 nm, with the photon energy close to the isomer energy. To obtain a laser beam with a suitable wavelength, it is proposed to use the seventh harmonic of the reference beam from an ytterbium-doped fiber laser with a wavelength of 1070 nm (Von der Wense, 2020). The reference beam is amplified as it passes through the optical resonator. In the same place, other harmonics, except for the seventh, are filtered out using the Brewster plate. Thus, a continuously generated beam with a power of 1.2 mW is obtained. Further in its path, a rotating shutter is placed, which interrupts the beam. As a result, the Fourier spectrum of the transformed beam takes the form of a frequency comb near the target line, which consists of several hundred equidistant teeth (Fig. 1). The power of each tooth is 10 nW, the half-width is 490 Hz (2 × 10$^{-12}$ eV), the distance between the teeth is 77 MHz=3 × 10$^{-8}$ eV.

**Estimation of the required scanning time by the IC mechanism in neutral atoms**

The laser beam is focused on a target, which is a thin circle with a diameter of 0.3 mm. 1.6×10$^{13}$ $^{229}$Th atoms are deposited on its surface by sputtering or otherwise. The scan duration of each pulse at a given frequency is 100 μs. During this time, approximately 15 atoms transfer into the isomeric state if the irradiation frequency falls into resonance with the nuclear transition. Isomeric nuclei decay into the ground state by internal conversion. Conversion electrons are registered using a multichannel counter. This signal serves as an indication that resonance has been reached. Otherwise, the teeth move further along the frequency scale, scanning the specified interval.

We emphasize that the use of IC is a decisive factor in this project. In the absence of an IC channel, the spectral width of each tooth should be of the order of the nuclear width, i.e., $10^{-19}$ eV, otherwise the isomer excitation cross section will be smaller in relation to the widths. The broadening of the nuclear line due to the IC makes it possible to use a comb with a spectral width of teeth up to $10^{-10}$ eV.

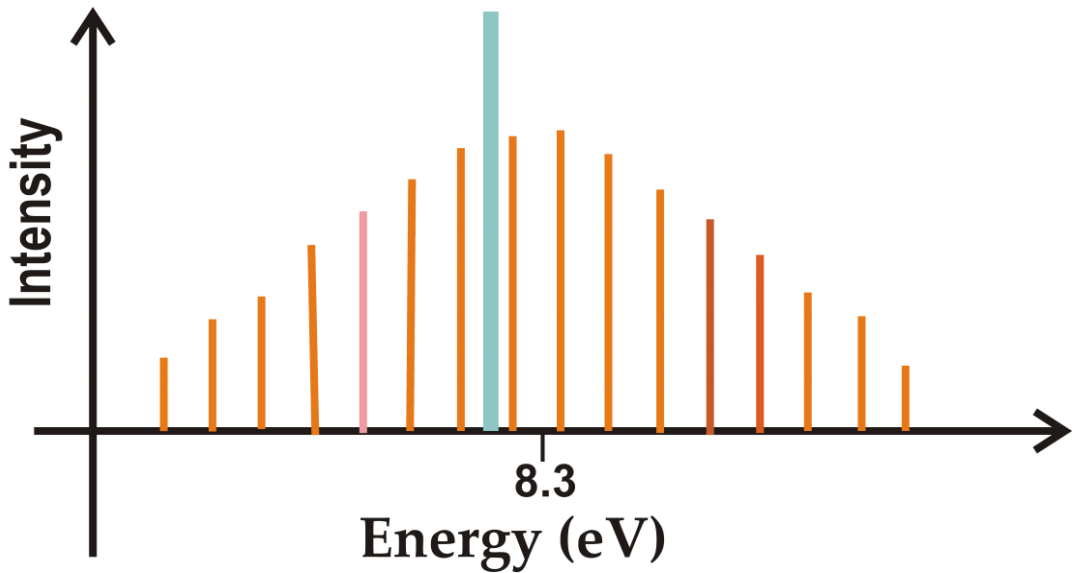

Fig. 1. Frequency comb proposed for isomer energy determination (Von der Wense, 2020). It consists of $1.2 \times 10^5$ comb modes (teeth) with frequencies, centered around 8.3 eV. The width of each mode is 490 Hz, and the power is 10 nW at the center of the comb. The position of the nuclear line is conditionally shown by a wide vertical segment. In the process of scanning the frequency of the comb teeth, a search is made for resonance with the nuclear frequency.

To evaluate the effectiveness of the scheme, we estimate the time required to scan the interval between the teeth of the comb. We set the frequency shift during scanning to be equal to the line width, that is, $10^{-10}$ eV. Then it will take approximately 5000 steps to scan. If one changes the comb frequency every second, one scan cycle will therefore take 5000 s. For a more accurate determination of the energy of the isomer, it is necessary to change the interval between the teeth. Therefore, several scan cycles may be required.

**Resonance pumping of the isomer through the electron shell**

The purpose of this work is to draw attention once again to the fact that it is possibility of significant reducing the time of the experiment if the resonant properties of the electron shell are used in full scale in order to enhance the effect of an external field on the nucleus. The example discussed above already uses the amplification of the isomeric transition by means of the IC channel by no less than 9 orders of magnitude. However, this can be said to be a kinematic gain. The line broadening resulting from IC is used to reduce the scan time. Strictly speaking, resonance with the electron shell as such is not exploited.

This method is based on the use of the properties of IC, due to which the broadening of the nuclear line allows the use of a comb with a spectral width of teeth up to $10^{-10}$ eV. Therefore, it is suitable for neutral atoms. Already in single ions, the ionization energy becomes greater than the transition energy, which turns off the VC channel. At the same time, most projects involve the use of $^{229}$Th ions.

But in the absence of the IC channel, the spectral width of each tooth should be of the order of the nuclear width, i.e., $10^{-19}$ eV, otherwise the isomer excitation cross section will be smaller by a factor of the width ratio. The dynamic broadening mechanism suitable for ions was considered by Karpeshin 1999. Let us consider its application on the example of single $^{229}$Th ions.

The essence of the method is illustrated by the Feynman graph in Fig. 2. The laser photon ω is absorbed by the 7$s$ electron, which passes into the virtual state. Near resonance, the 8$p$ electron makes the main contribution. It transfers part of the received energy to the nucleus, transferring it to the isomeric state. The electron remains in the excited 7$p$ state with the energy of $\omega_{7p}$. The resonance condition is that $\omega = \omega_n + \omega_{7p}$.

We note the features of the resonant excitation of the nucleus according to Fig. 2 compared with the absorption of a photon by a bare nucleus.

1. The cross section due to the resonance mechanism, according to the calculation (Karpeshin 1999), turns out to be enhanced by a factor of 40. Actually, this manifests the dynamic properties of the electron shell as a resonator.

2. The most important difference for the purposes of this work is the width of the resonance. If for a bare nucleus this width is given by the natural width (in the absence of IC) of the isomeric level of the nucleus $\Gamma_n$, then according to the mechanism of Fig. 2 it is equal to the sum of all the widths, electronic and nuclear ones. Usually, the sum of the widths of the intermediate 8$p$ and final 7$p$ atomic states dominates. Let us denote it as before by $\Gamma_a$. There is a relation $\Gamma_a \ggg \Gamma_n$, so scanning according to the mechanism of Fig. 2 requires by a factor of $\Gamma_a/\Gamma_n$ times less time.

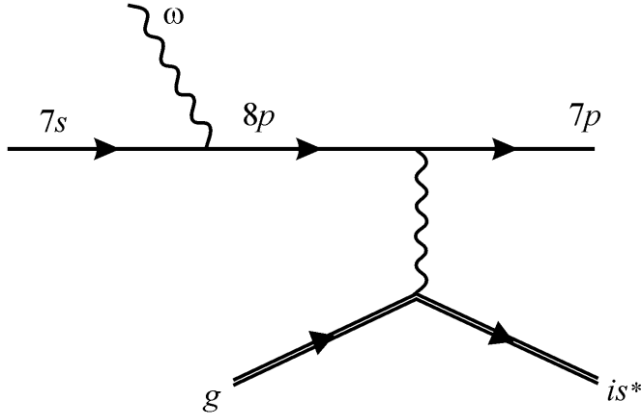

Fig. 2. Feynman graph of the resonance optical pumping the isomer.

For the effective realization of this mechanism, it is extremely important that the final atomic level, in the case, $7p$, should not be ground, but would have the usual atomic width. The virtual photon energy $\omega_i$ is, of course, equal to the isomer energy. But conservation of energy is ensured by the fact that the energy of the $7p$ state is automatically chosen on the profile of the $7p$ level. If there were a ground level instead of the $7p$ level, then the resonance would have a nuclear width $\Gamma_n$ in the absence of an atomic width $\Gamma_a$. The gain factor $R = 40$ would hold in the case of a monochromatic laser beam of photons with energy $\omega = \omega_i = \omega_n$, but the most important property for pumping the isomer with a spectral width of the laser beam as great as the atomic width would be lost. This moment was not taken into account in paper Porsev 2010. This and other shortcomings of the Ref. Porsev 2010 are discussed in detail in the papers by Karpeshin 2018, 2021.

**Scan time estimation using resonance properties of the electron shell**

Using the features of the Feynman diagram in Fig. 2, one can qualitatively estimate the gain in scanning time associated with its realization. Nuclear width is $10^{-10}$ eV. For the atomic width, we use a typical estimate $\Gamma_a \approx 10^{-8}$ eV, that is by a factor of 100 more than in the case of Th I. Crudely, this means that 100 teeth in Fig. 1 could be merged into one wide tooth, its power being by a factor of 100 greater than power of a single narrow tooth in Fig. 1. Furthermore, the broadening of the resonance provides a gain in scanning time by a factor of $\Gamma_a/\Gamma_n$, which grows up also by the same factor of 100 in comparison with the case of Th I. Finally, remembering the dynamic amplification of the diagram in Fig. 2, we obtain a gain in the scanning time required to establish the resonance as by a factor of $K = 400000$ times.

**Two mechanisms of optical pumping the isomers**

Let us consider qualitatively in more detail why the width of the intermediate electronic state is so important. Note that technically it may be easier to excite the input atomic state by two- or three-photon absorption, since the energy of one photon may not be enough. Although the project of single-photon pumping of the isomer by the optical comb method is actually considered above, it uses the seventh harmonic of the laser. To be definite, let us consider a typical mechanism of the two-photon partly resonance absorption in Th II, as considered in Karpeshin 2022. Its scheme is shown in Fig. 3.

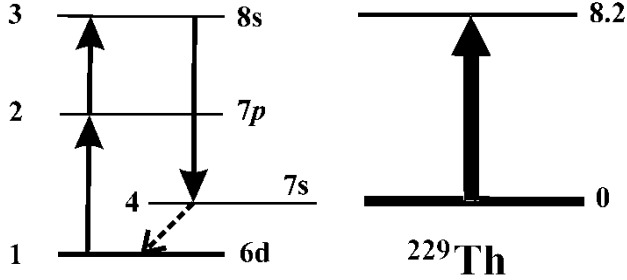

Fig. 3. Scheme of two-photon optical pumping.

Ground state 1 is described by configuration $7s6d^2$. State 2 corresponds to the $7s6d7p$ configuration with the energy of 3.084 eV. It is this transition that is characterized by the most intense line in the absorption spectrum. Therefore, the first photon is absorbed resonantly. Aa a result, the atom goes from the ground state 1 to excited state 2, the photon energy being equal to that of the atomic transition. Then the atom accumulates enough energy by absorbing the second photon and making the 2→3 transition. From the condition of the electric dipole absorption of the second quantum, it follows that state 3 can be either $8s$ or $7d$. It is also known from the IC theory that the probability of the next conversion transition $M1$ is directly proportional to the density of the squared electron wave functions on the nucleus. Therefore, this transition is the strongest if the electron passes between $s$-states. This condition rules out the $7d$ state, leaving the only possibility of $8s$. Hence state 3 should be close to the $7s8s6d$ level. The transition $7d→7s$ is at least three orders of magnitude weaker than $8s→7s$.

At the final stage, the atom transfers part of the absorbed energy to the nucleus and makes a transition into a state 4, remaining in the excited state. The energy of the 3→4 atomic transition is exactly equal to the energy of the isomer in accordance with the conservation of energy. Therefore, state 3 cannot be a real atomic state on the mass shell, since an exact resonance is impossible. Thus, the second photon is absorbed nonresonancely. This transition is inverse with respect to the IC process: the atom, being deexcited, transfers energy to the nucleus. Based on the IC theory, one can indicate the optimal quantum Th II numbers of the atomic states.

We note that from the experimental point of view, it is easier to excite a real 8s state with a second photon, since its energy is known. But then the resonance transition 3 → 4 becomes impossible: the nucleus must receive exactly the energy that is necessary to excite the isomer. Therefore, the 3 → 4 transition be-comes non-resonant, and the resulting smallness "eats" the gain from resonance upon absorption of the second photon. Accordingly, its probability is proportional to $\Gamma_4$, the width of level 4. The energy conservation law could be restored as a result of the next transition, for example, to the ground state $7s6d^2$. But this is a quadrupole $E2$ transition $7s\rightarrow 6d$, hindered by 4–6 orders of magnitude compared to the electric dipole $E1$. Thus, by no means could the state 4 be the main state in the optimal variant, and preference should be given to partially resonant pumping based on the reverse conversion.

The above qualitative considerations are fully confirmed by calculations. For the convenience of classification, the second mechanism was conditionally associated with the NEET processes, named after the paper by Morita, 1973, who proposed a mechanism for radiationless excitation of the nucleus, initiated by the creation of a hole in the inner shell. The paper by Karpeshin, 2017 shows that the ratio of the probabilities of the two mechanisms is:

$$\frac{W_{RC}}{W_{NEET}} = \frac{\Gamma_3}{\Gamma_4} \gg 1.$$

That says it all. The reverse conversion mechanism is more likely in proportion to the width ratio. Already from the energy dependence it is expected that $\Gamma_3 \gg \Gamma_4$. There are no other fundamental sources of the smallness. In the case of $^{229}$Th the inequality achieves as much as four orders of magnitude or more due to the quadrupole nature of the vertex $\Gamma_4$. This is not at all an accidental circumstance: there are simply no other states below. It's like the absence of excited E1 levels in atomic nuclei near the ground state. And if state 4 were the ground state, then its width should be set to zero: $\Gamma_4 = 0$. In this case, the probability of NEET excitation by the resonance mechanism would also be zero. Failure to take this argument into account in the work Porsev, 2010 is another fatal reason for the impossibility of practical implementation of the method proposed in it. Such is the true solution to the thorium puzzle (Karpeshin, 2021).

**Conclusion**

We are at the dawn of the nascence of new laser-nuclear technologies. They will be based on the resonance interactions of light beams with nuclei: resonance absorption and scattering, elastic and inelastic, nonlinear effects, such as the generation of higher harmonics, and others. In order to carry out these manipulations with bare nuclei, it is necessary to use spectroscopically narrow, almost monochromatic beams, with a spectral width within the nuclear linewidth. The development of such technologies requires knowledge of the energies of nuclear transitions and isomers. It is these beams that will form the basis of future nuclear-optical clocks.

At the same time, the electron shell can be used as an effective resonator that enhances the effect of light on the nucleus. The tool for designing such a resonator is provided by internal and, particularly, resonance conversion. The above examples once again demonstrate how efficiently resonance can be used for optical pumping of the $^{229}$Th isomeric state. Resonance can be used in two dimensions: either through amplification of the laser action, or weakening the requirements for the monochromaticity of the light beam. In the above example with Th I optical pumping, it is possible to obtain a gain of nine orders of magnitude in line broadening due to IC. In thorium ions, the resonance can be realized, which will give a combined gain, both in the strength of interaction - by a factor of 4000, and in broadening of the resonance line by another two orders of magnitude as compared to IC in neutral atoms.

**Gratitude**

The authors are grateful to Lars von der Wense for fruitful discussions.